\documentclass[conference,breaklinks=true,hyphens]{IEEEtran}
\IEEEoverridecommandlockouts

\usepackage{amsmath}
\usepackage{amssymb}
\usepackage{amsfonts}
\usepackage{algorithmic}
\usepackage{hyperref}
\usepackage{textcomp}
\usepackage{xcolor}

\usepackage{listings}
\usepackage{array}
\usepackage{booktabs}

\UseRawInputEncoding

\usepackage[labelfont=bf]{caption}

\usepackage{epstopdf}
\usepackage{comment}
\usepackage{textcase}

\usepackage{pgfplots}
\usepackage{pgfplotstable}
\pgfplotsset{compat=newest}
\pgfplotsset{compat=1.18}
\usetikzlibrary{3d}
\usepackage{eqparbox}
\usepackage{hyperref}  
\makeatletter
\g@addto@macro{\UrlBreaks}{\UrlOrds}
\makeatother
\usepackage{caption}
\usepackage{tikz}
\usepackage{tikz-3dplot}
\usetikzlibrary{positioning}
\usetikzlibrary{shapes.geometric, calc}
\usetikzlibrary{shapes.geometric, arrows}
\usetikzlibrary{shapes.geometric, arrows.meta}

\usetikzlibrary{positioning, shapes.geometric, arrows}
\tikzstyle{block} = [rectangle, rounded corners, minimum width=2.2cm, minimum height=1cm,text centered, draw=black, fill=cyan!30]
\tikzstyle{arrow} = [thick,->,>=stealth, color=red]

\tikzstyle{block} = [rectangle, rounded corners, minimum width=1.4cm, minimum height=0.6cm, text centered, draw=black, fill=cyan!30]
\tikzstyle{arrow} = [thick, ->, >=stealth, color=blue]

\usepackage{listings}

\tikzstyle{process} = [rectangle, rounded corners, minimum width=3cm, minimum height=1cm, text centered, draw=black, fill=blue!20]
\tikzstyle{arrow} = [thick,->,>=stealth]
\tikzstyle{process} = [rectangle, rounded corners, minimum width=2.5cm, minimum height=1cm, text centered, draw=black, fill=blue!20, text width=2.5cm]
\tikzstyle{arrow} = [thick,->,>=stealth]
\tikzstyle{process} = [rectangle, rounded corners, minimum width=2.5cm, minimum height=1cm, text centered, draw=black, fill=blue!20, text width=2.5cm]
\tikzstyle{arrow} = [thick,->,>=stealth]

\begin{document}
\title{Securing the Future of IVR: AI-Driven Innovation with Agile Security, Data Regulation, and Ethical AI Integration}

\author{
\IEEEauthorblockN{Khushbu Mehboob Shaikh}
\IEEEauthorblockA{\textit{Technical Lead, Principal Technical Account Manager} \\
\textit{Twilio Inc.}\\
Irving, Texas, United States \\
ORCID: 0009-0000-8681-5830
}
\and
\IEEEauthorblockN{Georgios Giannakopoulos}
\IEEEauthorblockA{\textit{Principal Engineer, Independent Researcher} \\
The Hague, The Netherlands \\
ORCID: 0000-0002-3707-3276}
}
\maketitle
\begin{abstract}
The rapid digitalization of communication systems has elevated Interactive Voice Response (IVR) technologies to become critical interfaces for customer engagement. With Artificial Intelligence (AI) now driving these platforms, ensuring secure, compliant, and ethically designed development practices is more imperative than ever. AI-powered IVRs leverage Natural Language Processing (NLP) and Machine Learning (ML) to personalize interactions, automate service delivery, and optimize user experiences. However, these innovations expose systems to heightened risks, including data privacy breaches, AI decision opacity, and model security vulnerabilities.

This paper analyzes the evolution of IVRs from static code-based designs to adaptive AI-driven systems, presenting a cybersecurity-centric perspective. We propose a practical governance framework that embeds agile security principles, compliance with global data legislation, and user-centric ethics. Emphasizing privacy-by-design, adaptive risk modeling, and transparency, the paper argues that ethical AI integration is not a feature but a strategic imperative. Through this multidimensional lens, we highlight how modern IVRs can transition from communication tools to intelligent, secure, and accountable digital frontlines—resilient against emerging threats and aligned with societal expectations.
\end{abstract}

\begin{IEEEkeywords}
Interactive Voice Response, IVR, Artificial Intelligence, AI, Natural Language Processing, NLP, Machine Learning, ML, AI-Powered Automation, Ethical AI Integration, Agile Security and Privacy, Information Security, Data Regulation, Privacy, Risk Analyses, Security Measures, Security Standards, Data Governance, Compliance, Trust, Cybersecurity Framework, User Perspective, Responsible Data Innovation, Strategy, Legislation on Security and Data Usage, Social Responsibility, Digital Transformation
\end{IEEEkeywords}

\section{Introduction}
Interactive Voice Response (IVR) systems have long served as essential digital entry points in customer service operations, enabling organizations to automate call handling, reduce wait times, and streamline user interactions \cite{Shaikh2024}. Traditionally, IVRs were constructed using rigid, code-heavy processes that required specialized technical expertise and offered limited adaptability. With the global push toward digital transformation, IVRs have evolved into more scalable, accessible, and intelligent platforms \cite{Teneo2024}.

The integration of Artificial Intelligence (AI) technologies—particularly Natural Language Processing (NLP) and Machine Learning (ML)—has profoundly redefined the function and potential of IVRs \cite{Coskun2019}. These modern AI-enhanced systems can interpret intent, personalize responses, and continuously adapt to user behavior, elevating user experience while improving operational efficiency. Yet, these innovations are accompanied by heightened risks related to data privacy, algorithmic transparency, and ethical governance \cite{eu2019ethics}.

In response to the autonomous nature of AI-driven platforms, organizations must adopt proactive strategies centered on agile security models, privacy-by-design, and strict adherence to regulations such as GDPR \cite{eu2016gdpr} and CCPA \cite{ccpa2020}. These measures should be embedded at the outset of development to mitigate evolving threats and ensure long-term system resilience. Ethical AI integration, meanwhile, is not merely a matter of technical compliance but a strategic imperative aligned with values of trust, transparency, and social responsibility.

In this paper, we present a multidimensional analysis of IVR development—from traditional architectures to contemporary AI-powered systems—through the lenses of cybersecurity, regulation, and ethics. We aim to inform a forward-looking, technically sound, and ethically governed approach to building intelligent IVR systems that meet today's dynamic operational and regulatory challenges \cite{Telnyx2024}, \cite{Cavoukian2009}. 

This paper explores the evolution of IVR systems through a cybersecurity and governance lens, offering insight into how organizations can design intelligent, secure, and ethically aligned IVR platforms that meet today's regulatory, technical, and social demands \cite{openai2023gpt4}.

\section{Traditional IVR Development: Complexity Without Control}
Legacy Interactive Voice Response (IVR) systems were traditionally developed using manually coded scripts and logic trees [1]. This method, while functional, demanded significant technical expertise and was inherently inflexible. Developers had to anticipate and define all possible call scenarios, embedding static routing rules, menu options, and voice prompts into large codebases. Any change in business logic or service offerings required full redeployment, often accompanied by rigorous retesting and extended downtimes.

\subsection{Limited Agility and High Maintenance Overhead}
These traditional systems lacked the agility needed in today’s fast-paced, regulation-driven environment \cite{Telnyx2024}. Because updates required direct code manipulation, response times to business needs or policy changes were slow. The absence of modular design and user-friendly interfaces made these systems inaccessible to non-technical stakeholders, creating communication gaps between operational teams and technical staff \cite{Shaikh2024}.

This siloed approach often excluded legal and compliance experts from early-stage development, increasing the risk of non-compliance with emerging legislation on data protection and information security \cite{openai2023gpt4}, \cite{weber2010iot}. Moreover, traditional IVRs did not implement risk analyses or privacy-by-design principles, making them ill-equipped to handle evolving privacy standards such as GDPR \cite{meta2022responsibleai}.

\subsection{Security and Compliance Limitations}
Security protocols in early IVR systems were typically reactive \cite{weber2010iot}. Basic authentication mechanisms, a lack of role-based access control, and minimal encryption made these systems vulnerable to threats such as spoofing, unauthorized access, and data interception \cite{Coskun2019}. Without clear governance and documentation, it was difficult to audit decision-making, identify vulnerabilities, or trace access to sensitive customer data \cite{Cavoukian2009}.

The inability to implement adaptive security measures meant that traditional IVRs often operated outside modern security standards \cite{Telnyx2024}. This not only posed reputational risks but also hindered scalability and integration with newer technologies that demand secure data interoperability \cite{ibm2023responsible}.

While traditional IVRs laid the foundation for automated communication, their inherent complexity, lack of built-in governance, and poor adaptability to current security and compliance requirements highlighted the need for more secure, user-focused, and agile IVR development approaches \cite{Voigt2017gdpr}.

\section{Widget-Based Platforms: Improved Access, New Risks}
To address the complexity of traditional IVR systems, the industry shifted toward widget-based development platforms. These tools introduced graphical user interfaces (GUIs) \cite{Twilio2023} that enabled users to visually design call flows using pre-built components, or ``widgets.'' By eliminating the need for manual coding, widget-based platforms democratized IVR development, making it accessible to non-technical users such as service managers and business analysts \cite{Coskun2019}, \cite{Coskun-Setirek2019}.

\subsection{Democratization and Speed of Deployment}
With drag-and-drop functionality and reusable modules, organizations could quickly prototype, iterate, and deploy IVR solutions in response to changing business needs \cite{Goodman2017right}, \cite{nist2018csf}. This low-code environment significantly reduced development cycles and improved responsiveness to market demands. Additionally, the visual nature of these platforms encouraged cross-functional collaboration, allowing operational and legal teams to be more actively involved in the design of customer interactions \cite{ibm2023responsible}.

This acceleration, however, came with its own set of challenges \cite{Gonzalez2016pbd}. The ease of access often led to inconsistent security implementations and misconfigured privacy settings \cite{weber2010iot}. Without built-in checks for regulatory compliance or enforced security standards, these platforms risked exposing sensitive customer data—especially when integrated with external systems \cite{iso2022}.

\subsection{Security Risks and Governance Gaps} 
While widget-based tools enabled agility, they also introduced new attack vectors \cite{Coskun-Setirek2019}. IVR flows designed by untrained personnel may lack proper authentication, encryption, or logging---leaving systems vulnerable to threats like voice phishing (vishing), session hijacking, or unauthorized data access. Moreover, insufficient governance in managing changes and version control could result in compliance failures and audit deficiencies \cite{Mittelstadt2016ethics}.

To mitigate these risks, organizations must embed agile security and privacy practices into the platform itself \cite{weber2010iot}. Role-based access control, compliance checklists, audit logging, and risk analysis workflows should be standard components of any widget-based IVR system \cite{salesforce2023ethicalai}. Establishing security policies and continuous monitoring processes ensures that the convenience of visual development does not compromise regulatory adherence or trust \cite{Goodman2017right}.

Ultimately, while widget-based platforms represent progress in digitalization, they must be governed carefully to support secure, privacy-conscious, and ethically sound IVR deployment \cite{openai2023gpt4}.

\begin{figure}[ht]
    \centering
    \includegraphics[width=0.5\textwidth]{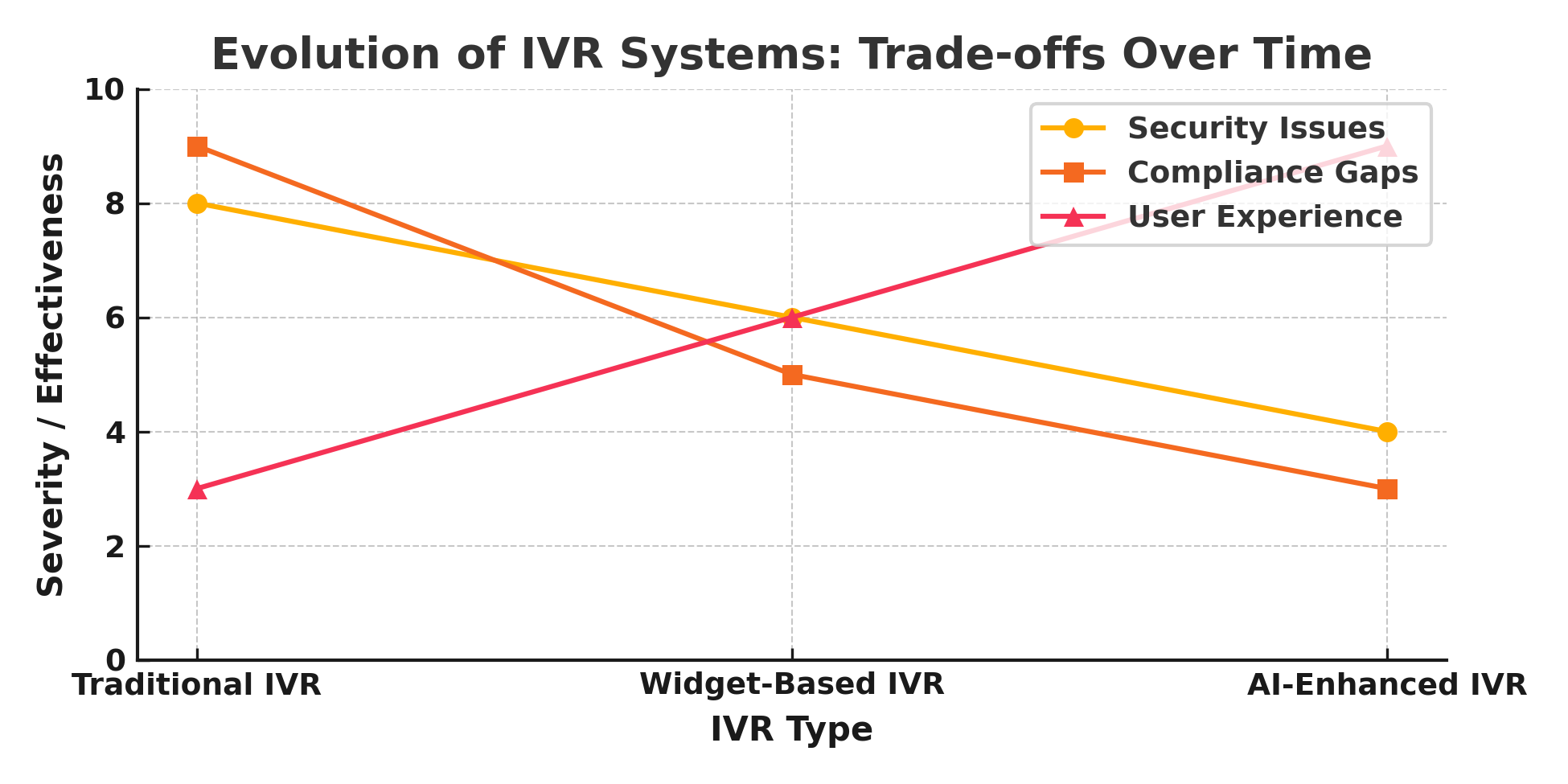} 
    \caption{Evolution of IVR Systems: Trade-offs Over Time.}
    \label{fig1}
\end{figure}

\hyperref[fig1]{Figure 1} illustrates the evolution of IVR systems, showing a clear trade-off between security risks, compliance gaps, and user experience across three generational stages: Traditional, Widget-Based, and AI-Enhanced IVRs. As seen, traditional IVRs are plagued by high security and compliance concerns but offer limited user satisfaction. Widget-based systems improve user experience moderately but still present governance challenges. AI-enhanced IVRs, while offering the highest user experience, introduce new forms of security threats, although these are more manageable under agile and ethical frameworks. This trend highlights the necessity of embedding robust cybersecurity and ethical AI practices to maintain trust and compliance in modern IVR systems.

\section{The Role of AI in Modern IVR Systems}
Artificial Intelligence (AI) technologies, particularly Natural Language Processing (NLP) and Machine Learning (ML), are transforming Interactive Voice Response (IVR) systems into intelligent, context-aware platforms \cite{Prabha2017}, \cite{Telnyx2023}. These systems now understand user intent, adapt to behaviors, and personalize responses in real time, greatly improving operational efficiency and user satisfaction \cite{Teneo2024}, \cite{ibm2023responsible}, \cite{google2023dialogflow}.

\subsection{Automation and Personalization} 
AI enables IVR systems to go beyond menu trees, allowing users to speak naturally and receive dynamic, context-sensitive responses \cite{Cavoukian2009}. These systems learn from historical interactions to optimize routing, anticipate questions, and reduce wait times \cite{Mctear2016}. As a result, they provide a more intuitive and satisfying experience for the user while minimizing strain on human agents \cite{Jurafsky2023}, \cite{HIPAA2024}.

However, increased automation requires responsible oversight \cite{Coskun-Setirek2019}. The collection, analysis, and storage of user voice data demand rigorous adherence to privacy laws and ethical handling. AI models should be trained and deployed under strict data regulation frameworks to avoid misuse or unintended bias \cite{Coskun2019}, \cite{obermeyer2019bias}.

\subsection{Ethics, Transparency, and User Trust} 
AI-powered IVRs often operate as black-box systems, making it difficult to trace decision-making logic. This opacity can erode user trust, particularly if errors or unintended actions occur \cite{Shaikh2024}, \cite{Coskun2019}. To mitigate this, organizations must adopt explainable AI (XAI) principles, ensuring that decisions made by the system are auditable and comprehensible to both developers and regulators \cite{Voigt2017gdpr}, \cite{Coskun-Setirek2019}.

Embedding ethical AI integration involves considering consent, fairness, and transparency throughout the system’s lifecycle \cite{Shaikh2024}, \cite{Mctear2016}. It also means creating escalation paths where human agents intervene in high-stakes or sensitive situations \cite{Coskun-Setirek2019}. Governance structures should clearly define accountability for AI outcomes and include risk analyses to manage emerging threats \cite{Gonzalez2016pbd}.

\subsection{Security Integration}  
With the advent of AI, IVR systems have transitioned into complex ecosystems with expanded attack surfaces \cite{Coskun-Setirek2019}. New vulnerabilities such as voice spoofing, adversarial inputs, and model poisoning require a recalibrated security posture. It is no longer sufficient to retrofit traditional controls into AI systems; instead, a paradigm shift toward proactive and adaptive defense mechanisms is essential \cite{eu2019ethics}, \cite{weber2010iot}, \cite{Goodman2017right}. To ensure secure deployment, AI-enhanced IVRs must implement stringent safeguards such as encrypted voice channels, input validation layers, model audit trails, and adversarial testing environments. AI must be managed as a critical infrastructure component—subject to the same, if not higher, levels of scrutiny as conventional systems \cite{eu2016gdpr}, \cite{Telnyx2024}, \cite{nist2018csf}.

In alignment with standards such as ISO/IEC 27001 and the NIST Cybersecurity Framework \cite{nist2018csf}, organizations must integrate real-time monitoring, anomaly detection, and continuous penetration testing into the AI-IVR lifecycle. Furthermore, cross-functional security design reviews should be mandatory for all releases, ensuring that evolving ML models do not inadvertently introduce new threats \cite{ibm2023responsible}, \cite{iso2022}, \cite{google2023dialogflow}.

Ultimately, AI should not be perceived as a plug-and-play enhancement, but as a sophisticated subsystem that mandates a bespoke security framework encompassing both traditional cybersecurity best practices and emerging AI-specific controls.

{\fontsize{9pt}{10pt}\selectfont
\begin{table}[htbp]
\centering
\footnotesize
\begin{tabular}{llll}
\toprule
\textbf{Aspect} & \textbf{Traditional} & \textbf{Widget-Based} & \textbf{AI-Enhanced} \\
\midrule
Dev. Approach & Manual coding & GUI-based & AI/NLP logic \\
Technical Expertise & High & Moderate & Mod-High \\
User Accessibility & Low & High & Very High \\
Security Controls & Reactive & Inconsistent & Proactive \\
Compliance Level & Limited & Partial & Strong, Audited \\
User Experience & Low & Moderate & Personalized \\
Update Agility & Slow & Faster & Continuous \\
Explainability & None & Limited & Built-in XAI \\
Integration Readiness & Low & Medium & High (with APIs) \\
\bottomrule
\end{tabular}
\caption{Comparison of IVR System Generations}
\label{tab1}
\end{table}
}

\hyperref[tab1]{Table I} summarizes the evolution of IVR systems across technical and operational metrics. Traditional IVRs are rigid and developer-dependent. Widget-based platforms increase agility but require governance controls. AI-enhanced IVRs deliver the highest personalization and demand advanced oversight mechanisms.

\section{Cybersecurity and Compliance Framework}
As AI-enhanced IVR systems become integral to customer service infrastructure, organizations must adopt a holistic cybersecurity and compliance framework \cite{Teneo2024}, \cite{Telnyx2023}, \cite{HIPAA2024}. This framework must ensure alignment with legal obligations, industry-specific security standards, and evolving user expectations \cite{weber2010iot}. Without proper controls, even advanced IVRs can become vectors for data breaches, non-compliance, and reputational damage \cite{meta2022responsibleai}, \cite{cisco2024ai},  \cite{accenture2022ethicalai}.

\subsection{Legislation on Security, Privacy, and Data Usage}   
Modern IVR systems operate within a complex legal ecosystem that includes global data protection regulations such as the General Data Protection Regulation (GDPR) \cite{eu2019ethics}, \cite{eu2016gdpr}, \cite{Goodman2017right}, California Consumer Privacy Act (CCPA) \cite{ccpa2020}, and regional telecom and financial compliance rules \cite{weber2010iot}. These laws emphasize principles such as data minimization, user consent, transparency, and the right to be forgotten \cite{Gonzalez2016pbd}, \cite{Mittelstadt2016ethics}.

AI-powered IVRs, which often record, analyze, and store voice interactions, must therefore implement privacy-by-design strategies \cite{weber2010iot}, \cite{gartner2023ivr}. This includes anonymization techniques, informed consent flows, data retention policies, and clear mechanisms for users to exercise their data rights \cite{weber2010iot}. Failing to meet these obligations can result in regulatory fines and erosion of customer trust \cite{Telnyx2024}, \cite{openai2023gpt4}, \cite{Banham2018}.

\subsection{Security Standards and Risk Analysis}  
Compliance alone does not ensure security \cite{weber2010iot}. Organizations must align their IVR systems with established cybersecurity standards such as ISO/IEC 27001 \cite{iso2022}, NIST Cybersecurity Framework \cite{nist2018csf}, and OWASP guidelines \cite{Owasp2010}, \cite{OWASPAI2025}, \cite{Clinton2025}. These standards provide a foundation for implementing threat detection, access control, incident response, and business continuity planning \cite{Hao2019}.

Risk analyses must be conducted throughout the system’s life-—cycle from design and development to deployment and decommissioning \cite{Coskun-Setirek2019}. This involves identifying attack surfaces, evaluating system vulnerabilities, and implementing controls like end-to-end encryption, multi-factor authentication, and role-based access management \cite{meta2022responsibleai}, \cite{nist2018csf}, \cite{cisco2024ai}.

Regular vulnerability assessments and penetration testing are necessary to validate the security posture of the IVR system. Logging and monitoring, supported by real-time analytics, ensure quick detection of anomalies or malicious activity, thereby reducing mean time to response (MTTR) \cite{Shaikh2024}, \cite{Twilio2023}, \cite{Owasp2010}.

\subsection{Agile and Adaptive Security Strategy}   
In the face of constantly evolving threats, a static security model is insufficient. IVR systems must adopt agile security practices that allow continuous integration of new safeguards without disrupting operations \cite{Voigt2017gdpr}, \cite{Gonzalez2016pbd}. This includes automated security testing during development, continuous compliance checks, and modular policy updates.

Security and compliance must not be perceived as bottlenecks but as enablers of innovation \cite{Coskun-Setirek2019}. Agile approaches support iterative development and allow organizations to adapt swiftly to regulatory changes or identified vulnerabilities \cite{weber2010iot}. Cross-functional teams-—including security engineers, compliance officers, and legal advisors-—must collaborate from the earliest design phase to ensure a shared understanding of responsibilities \cite{Coskun2019}, \cite{Telnyx2024}, \cite{Mctear2016}.

An adaptive strategy also includes educating stakeholders about best practices in data handling, ethical AI use, and social responsibility. Training programs and internal policy audits help create a security-first culture that scales with technological advancements \cite{salesforce2023ethicalai}, \cite{cisco2024ai}, \cite{Hao2019}.

A robust cybersecurity and compliance framework is essential for building resilient, user-trusted, and legally compliant IVR systems \cite{Teneo2024}. By integrating legislation, security standards, and agile governance practices, organizations can create sustainable systems that meet the demands of modern digital ecosystems \cite{Telnyx2023}, \cite{HIPAA2024}.

\section{Ethical AI Integration: A Strategic Imperative}  
As AI becomes embedded in mission-critical systems like IVR platforms, ethical considerations are no longer optional--—they are strategic imperatives \cite{Telnyx2023}. Ethical AI integration ensures that technological advancements align with societal values, legal expectations, and user trust \cite{meta2022responsibleai}, \cite{Gonzalez2016pbd}, \cite{OWASPAI2025}. For IVR systems that interact with a wide demographic, including vulnerable populations, this means designing for fairness, inclusivity, and transparency from the outset \cite{Teneo2024}, \cite{HIPAA2024}.

\subsection{Privacy by Design and User Perspective}   
Ethical AI begins not with compliance checklists but with an unwavering focus on user-centricity \cite{Telnyx2023}. IVR systems, often the first point of contact between organizations and users, must embed privacy-by-design principles at the architectural level \cite{Teneo2024}. This includes not only minimizing data collection to what is strictly necessary, but also ensuring that consent mechanisms are granular, contextual, and auditable \cite{Voigt2017gdpr}, \cite{Gonzalez2016pbd}.

AI-IVRs must empower users by offering clear options to opt out of data retention, view explanations of automated decisions, and request human oversight where appropriate. Interfaces should be accessible across linguistic, cultural, and cognitive dimensions—ensuring inclusivity and minimizing bias \cite{Mctear2016}, \cite{Jurafsky2023}, \cite{obermeyer2019bias}.

Critically, ethical implementation requires the co-design of systems with diverse stakeholders, including ethicists, legal experts, and marginalized user groups. This participatory approach ensures that the deployment of AI technologies aligns not only with legal requirements but also with broader human values.

\subsection{Fairness, Transparency, and Accountability}  
AI systems that operate without oversight risk reinforcing existing biases or making opaque decisions that affect customer outcomes \cite{Coskun-Setirek2019}. Ethical AI integration demands transparency—-through documentation of model training \cite{Hao2019}, clear explanations of automated decisions, and auditability of system behavior \cite{ibm2023responsible}, \cite{google2023dialogflow}, \cite{gartner2023ivr}.

Accountability mechanisms must also be established. This includes defining who is responsible when the system fails, whether due to biased data, technical error, or lack of escalation \cite{Mittelstadt2016ethics}, \cite{obermeyer2019bias}. AI systems must be regularly evaluated for fairness across different demographic groups, and organizations should publish findings or conduct third-party assessments to reinforce credibility \cite{Coskun2019}, \cite{Mctear2016}.

\subsection{Social Responsibility and Governance}   
Beyond compliance, organizations must consider the social impact of their AI implementations \cite{meta2022responsibleai}, \cite{Banham2018}. Governance structures should be in place to evaluate ethical risks alongside business benefits \cite{HIPAA2024}. Ethical review boards, internal ethics officers, or cross-disciplinary working groups can guide responsible development and deployment practices \cite{eu2019ethics}, \cite{eu2016gdpr}, \cite{Voigt2017gdpr}, \cite{Goodman2017right}, \cite{Clinton2025}.

Ultimately, ethical AI integration in IVR is about aligning technical excellence with human values \cite{Coskun-Setirek2019}. By embedding governance, transparency, and empathy into AI systems, organizations foster trust and set a standard for responsible digital transformation \cite{ibm2023responsible}, \cite{salesforce2023ethicalai}.

\section{Challenges and Mitigation Approaches} 
The integration of AI into IVR systems introduces powerful capabilities but also surfaces several operational, legal, and ethical challenges \cite{iso2022}, \cite{salesforce2023ethicalai}. To fully leverage the benefits of intelligent automation while safeguarding privacy and user trust, organizations must proactively identify and mitigate emerging risks \cite{eu2019ethics}, \cite{eu2016gdpr}, \cite{ccpa2020}, \cite{nist2018csf}.

\begin{table}[htbp]
\centering
\begin{tabular}{p{2.5cm} p{5.0cm}}
\toprule
\textbf{Challenge} & \textbf{Mitigation Strategy} \\
\midrule
Privacy Risks & End-to-end encryption, data anonymization, privacy impact assessments (PIAs) \\
Legacy Compatibility & Middleware, phased rollout, API standardization \\
User Expectations & Disclosures, consent prompts, HITL design \\
Algorithmic Bias & Diverse data, XAI, fairness audits \\
Regulatory Compliance & GDPR/CCPA audits, lifecycle documentation \\
Governance Gaps & Ethics boards, defined roles, audit controls \\
\bottomrule
\end{tabular}
\caption{Challenges and Mitigations in AI-Driven IVRs}
\label{tab2}
\end{table}

\hyperref[tab2]{Table II} provides an overview of primary technical, legal, and ethical risks in deploying AI-driven IVRs, along with actionable mitigation strategies. It emphasizes the need to balance innovation with compliance and inclusivity to ensure resilient and trustworthy platforms.

As AI systems become increasingly autonomous, organizations must address data privacy risks, legacy system compatibility issues, and the necessity of ethical AI practices. Solutions such as explainable AI, privacy-by-design, and regulatory alignment are critical for maintaining trust, minimizing risk, and ensuring responsible innovation.

\subsection{Data Privacy and Security Risks} 
AI-powered IVRs process large volumes of sensitive user data, including personal identifiers and behavioral insights \cite{weber2010iot}. During model training and real-time interaction, improper data handling can lead to leakage, unauthorized access, or even model inversion attacks \cite{Mittelstadt2016ethics}, \cite{gartner2023ivr}. Additionally, datasets used to train AI may carry historical biases or lack consent-compliant provenance \cite{obermeyer2019bias}.

To mitigate these risks, organizations must enforce strong data governance policies: end-to-end encryption, strict access controls, anonymization of training datasets, and regular audits \cite{eu2019ethics}, \cite{Goodman2017right}. Privacy impact assessments (PIAs) should be conducted before deployment and during system updates, ensuring compliance with data regulation frameworks such as GDPR \cite{eu2016gdpr}, \cite{Goodman2017right}, HIPAA \cite{HIPAA2024}, or CCPA \cite{ccpa2020}.

\subsection{Legacy Integration and System Compatibility}  
Many organizations still operate legacy telephony and call center infrastructure that is not built to support AI functionality \cite{eu2016gdpr}, \cite{ccpa2020}, \cite{nist2018csf}, \cite{Owasp2010}, \cite{OWASPAI2025}. Integrating new systems with outdated frameworks increases the risk of configuration errors, security gaps, or poor performance \cite{weber2010iot}. Seamless integration requires middleware, API standardization, and well-defined data exchange protocols \cite{Coskun2019}, \cite{Cavoukian2009}, \cite{Voigt2017gdpr}.

Mitigation involves architectural reviews, interoperability testing, and phased deployment plans. Cybersecurity teams should collaborate with developers to evaluate legacy weaknesses and ensure that secure bridges are used to connect old and new components \cite{Prabha2017}, \cite{obermeyer2019bias}.

\subsection{User Expectation Management and Human Escalation} 
AI systems that mimic human communication can lead users to overestimate their capabilities \cite{HIPAA2024}. When IVRs fail to resolve issues or escalate appropriately, frustration and reputational damage can follow. Ethical concerns also arise when users are unaware they are speaking to an AI agent \cite{Mittelstadt2016ethics}, \cite{Mctear2016}, \cite{Jurafsky2023}.

Clear communication, such as AI disclosures, feedback prompts, and escalation paths to human agents, is essential. Human-in-the-loop design ensures that complex or emotionally sensitive cases are handled appropriately, maintaining empathy and care in customer interactions \cite{Shaikh2024}, \cite{Prabha2017}, \cite{Jurafsky2023}.

By addressing these challenges through structured compliance strategies, technical safeguards, and governance frameworks, organizations can build secure, trustworthy, and effective AI-enhanced IVR systems \cite{Teneo2024}.

\section{Conclusion}
Interactive Voice Response (IVR) systems are no longer merely automated call routing tools; they are evolving into complex, intelligent interfaces that embody the values and priorities of the organizations they serve \cite{meta2022responsibleai}, \cite{Twilio2023}, \cite{cisco2024ai}, \cite{accenture2022ethicalai}.  Their transformation—enabled by AI and Natural Language Processing (NLP)—promises substantial benefits in user experience, operational efficiency, and personalization. Yet, without comprehensive governance, this evolution could just as easily produce risk, distrust, and systemic vulnerability \cite{Telnyx2024}, \cite{openai2023gpt4}, \cite{ibm2023responsible}, \cite{iso2022}, \cite{salesforce2023ethicalai}, \cite{google2023dialogflow}, \cite{Banham2018}.

The integration of Artificial Intelligence (AI) and Natural Language Processing (NLP) has enabled IVRs to automate tasks, personalize interactions, and respond intelligently to diverse customer needs \cite{Owasp2010}, \cite{OWASPAI2025}, \cite{Clinton2025}. This paper presents a roadmap for secure, ethical, and regulation-aligned IVR system development. We advocate for an approach that treats cybersecurity and ethical AI as foundational—not ancillary—components of digital transformation \cite{eu2016gdpr}, \cite{ccpa2020}, \cite{nist2018csf}. 

Privacy-by-design, adaptive security practices, and explainable decision logic should be embedded in every layer of system architecture \cite{eu2019ethics}, \cite{Owasp2010}, \cite{OWASPAI2025}.

Future-ready IVRs must be governed by interdisciplinary teams that span technical, legal, ethical, and human-centered domains. These teams should be tasked with ensuring accountability, transparency, and inclusivity at every stage of the system lifecycle \cite{Voigt2017gdpr}, \cite{Prabha2017}, \cite{Mctear2016}. In particular, use cases in healthcare \cite{HIPAA2024}, finance, and government demand heightened scrutiny and continuous evaluation.

Organizations must recognize IVRs not just as communication tools but as digital frontlines that handle sensitive data and influence user perceptions \cite{Shaikh2024}, \cite{Mittelstadt2016ethics}, \cite{Jurafsky2023}. Governance structures must support explainability, fairness, and accountability, while adaptive security measures ensure resilience against evolving cyber threats \cite{Prabha2017}, \cite{obermeyer2019bias}. Equally important is the consideration of user perspective and social responsibility, ensuring that technology serves people equitably and transparently \cite{Coskun2019}, \cite{Cavoukian2009}.

As AI capabilities grow, so too must our commitment to responsible innovation. Ethical AI integration is not a checkbox; it is a mindset that must guide every engineering and policy decision \cite{Shaikh2024}, \cite{Goodman2017right}, \cite{Mittelstadt2016ethics}. As AI continues to evolve, organizations must invest in interdisciplinary collaboration, continuous monitoring, and stakeholder education to maintain both trust and performance \cite{Voigt2017gdpr}, \cite{Jurafsky2023}.

By embedding trust, resilience, and equity into the DNA of IVR platforms, organizations can lead with both technological excellence and moral clarity \cite{weber2010iot}, \cite{Twilio2023}, \cite{accenture2022ethicalai}.

\section*{Acknowledgment}
The authors would like to acknowledge the critical role of AI-powered research and analytics platforms in shaping the direction and insights presented in this paper. The ability to conduct cross-domain evaluations of ethical, technical, and legal frameworks was significantly enhanced through these technologies.

We also recognize the contributions of global standards organizations (e.g., ISO, NIST, OWASP) and academic institutions whose publications and case studies informed the framework and analysis proposed. Special thanks go to interdisciplinary experts from AI ethics, cybersecurity, and human-computer interaction communities, whose discussions inspired several of the principles described in this work.

Lastly, we extend our gratitude to the reviewers and professionals who continue to shape the responsible evolution of voice technologies, ensuring that innovation proceeds with equity, transparency, and resilience.

\section*{Additional information}

\noindent
The authors declare that they have no conflict of interest or competing interests.

\bibliographystyle{IEEEtran}
\bibliography{anystyle}

\begin{thebibliography}{10}
\providecommand{\url}[1]{#1}
\csname url@samestyle\endcsname
\providecommand{\newblock}{\relax}
\providecommand{\bibinfo}[2]{#2}
\providecommand{\BIBentrySTDinterwordspacing}{\spaceskip=0pt\relax}
\providecommand{\BIBentryALTinterwordstretchfactor}{4}
\providecommand{\BIBentryALTinterwordspacing}{\spaceskip=\fontdimen2\font plus
\BIBentryALTinterwordstretchfactor\fontdimen3\font minus \fontdimen4\font\relax}
\providecommand{\BIBforeignlanguage}[2]{{%
\expandafter\ifx\csname l@#1\endcsname\relax
\typeout{** WARNING: IEEEtran.bst: No hyphenation pattern has been}%
\typeout{** loaded for the language `#1'. Using the pattern for}%
\typeout{** the default language instead.}%
\else
\language=\csname l@#1\endcsname
\fi
#2}}
\providecommand{\BIBdecl}{\relax}
\BIBdecl

\bibitem{Shaikh2024}
K.~M. Shaikh and G.~Giannakopoulos, ``Evolution of ivr building techniques: from code writing to ai-powered automation,'' \emph{Computer Science. Software Engineering (cs.SE)}, no.~1, pp. 1--6, Nov 2024.

\bibitem{Teneo2024}
Teneo.ai, ``5 reasons why ignoring conversational ivr systems could cost your business big time,'' \url{https://www.teneo.ai/blog/ignoring-conversational-ivr-systems-could-cost-your-business}, Teneo.ai, Tech. Rep., Oct 2024.

\bibitem{Coskun2019}
M.~Coskun-Setirek and Z.~Tanrikulu, ``Evaluation of interactive voice response (ivr) systems from the users’ perspective,'' \emph{International Journal of Human–Computer Interaction}, vol.~35, no.~1, pp. 61--69, 2019.

\bibitem{eu2019ethics}
EC, ``Ethics guidelines for trustworthy ai,'' \url{https://ec.europa.eu/newsroom/dae/document.cfm?doc_id=60419}, European Commission High-Level Expert Group on AI, Tech. Rep., 2019.

\bibitem{eu2016gdpr}
\BIBentryALTinterwordspacing
EU, ``General data protection regulation (gdpr),'' European Union, Tech. Rep., 2016, regulation (EU) 2016/679. [Online]. Available: \url{https://eur-lex.europa.eu/eli/reg/2016/679/oj}
\BIBentrySTDinterwordspacing

\bibitem{ccpa2020}
SCDJ, ``California consumer privacy act (ccpa) and california privacy rights act (cpra),'' \url{https://oag.ca.gov/privacy/ccpa}, State of California Department of Justice, Tech. Rep., 2020.

\bibitem{Telnyx2024}
Telnyx, ``Conversational ai and its impact on ivr systems,'' \url{https://telnyx.com/resources/conversational-ai-ivr}, Telnyx, Tech. Rep., Oct 2024.

\bibitem{Cavoukian2009}
\BIBentryALTinterwordspacing
A.~Cavoukian, ``Privacy by design: The 7 foundational principles,'' Information and Privacy Commissioner of Ontario, Canada, Tech. Rep., 2009. [Online]. Available: \url{https://www.ipc.on.ca/wp-content/uploads/Resources/7foundationalprinciples.pdf}
\BIBentrySTDinterwordspacing

\bibitem{openai2023gpt4}
\BIBentryALTinterwordspacing
J.~Achiam, S.~Adler, S.~Agarwal, L.~Ahmad, I.~Akkaya, F.~Aleman, D.~Almeida, J.~Altenschmidt, S.~Altman, S.~Anadkat, R.~Avila, I.~Babuschkin, S.~Balaji, V.~Balcom, P.~Baltescu, H.~Bao, M.~Bavarian, J.~Belgum, and B.~Zoph, ``Gpt-4 technical report,'' OpenAI, Tech. Rep., 03 2023. [Online]. Available: \url{https://arxiv.org/abs/2303.08774}
\BIBentrySTDinterwordspacing

\bibitem{weber2010iot}
R.~H. Weber, ``Internet of things – new security and privacy challenges,'' \emph{Computer Law \& Security Review}, vol.~26, no.~1, pp. 23--30, 2010.

\bibitem{meta2022responsibleai}
MetaAI, ``Responsible ai practices at meta,'' \url{https://ai.facebook.com/responsible-ai/}, Meta AI, Tech. Rep., 2022.

\bibitem{ibm2023responsible}
IBM, ``Ai ethics: Ibm's ai ethics principles, pillars, practices, and policies,'' \url{https://ifhp.com/wp-content/uploads/2023/12/IBM-Data-Ethics-how-to-operationalize-MLAI-while-respecting-the-ethical-aspects.pdf}, IBM, Tech. Rep., Dec 2023.

\bibitem{Voigt2017gdpr}
P.~Voigt and A.~Von~dem Bussche, \emph{The EU General Data Protection Regulation (GDPR): A Practical Guide}.\hskip 1em plus 0.5em minus 0.4em\relax Springer, 2017.

\bibitem{Twilio2023}
Twilio, ``Getting started with twilio studio,'' \url{https://www.twilio.com/docs/studio}, Twilio, Tech. Rep., 2023.

\bibitem{Coskun-Setirek2019}
A.~Coskun-Setirek and Z.~Tanrikulu, ``Intelligent interactive voice response systems and customer satisfaction,'' \emph{International Journal of Advanced Trends in Computer Science and Engineering}, vol.~8, pp. 4--11, 03 2019.

\bibitem{Goodman2017right}
B.~Goodman and S.~Flaxman, ``European union regulations on algorithmic decision-making and a 'right to explanation','' \emph{AI Magazine}, vol.~38, no.~3, pp. 50--57, 2017.

\bibitem{nist2018csf}
\BIBentryALTinterwordspacing
NIST, ``Framework for improving critical infrastructure cybersecurity, version 1.1,'' U.S. Department of Commerce, Tech. Rep., 2018. [Online]. Available: \url{https://nvlpubs.nist.gov/nistpubs/CSWP/NIST.CSWP.04162018.pdf}
\BIBentrySTDinterwordspacing

\bibitem{Gonzalez2016pbd}
N.~Gonzalez \emph{et~al.}, ``A framework for the design and implementation of privacy by design,'' \emph{Information Systems Frontiers}, vol.~18, pp. 315--331, 2016.

\bibitem{iso2022}
ISO, ``Iso/iec 27001:2022 information security management systems,'' International Organization for Standardization, Tech. Rep., 2022.

\bibitem{Mittelstadt2016ethics}
B.~Mittelstadt, P.~Allo, M.~Taddeo, S.~Wachter, and L.~Floridi, ``The ethics of algorithms: Mapping the debate,'' \emph{Big Data \& Society}, vol.~3, no.~2, pp. 1--21, 2016.

\bibitem{salesforce2023ethicalai}
Salesforce, ``Ai ethics guidelines,'' \url{https://www.salesforce.com/company/privacy/ethical-ai/}, Salesforce, Tech. Rep., 2023.

\bibitem{Prabha2017}
M.~S. Prabha and P.~R. Deshmukh, ``Artificial intelligence based interactive voice response system,'' \emph{International Journal of Computer Applications}, vol. 174, no.~2, pp. 21--25, 2017.

\bibitem{Telnyx2023}
Telnyx, ``Improve your smart ivr and conversational ai,'' \url{https://telnyx.com/resources/number-lookup-smart-ivr-conersational-ai}, Telnyx, Tech. Rep., Sep 2023.

\bibitem{google2023dialogflow}
Google, ``Dialogflow cx: Next-gen ai conversational platform,'' \url{https://cloud.google.com/dialogflow/docs}, Google Cloud, Tech. Rep., 2023.

\bibitem{Mctear2016}
M.~McTear, Z.~Callejas, and D.~Griol, \emph{The Conversational Interface: Talking to Smart Devices}.\hskip 1em plus 0.5em minus 0.4em\relax Springer, 2016.

\bibitem{Jurafsky2023}
\BIBentryALTinterwordspacing
D.~Jurafsky and J.~H. Martin, \emph{Speech and Language Processing}, 3rd~ed., 2023. [Online]. Available: \url{https://web.stanford.edu/~jurafsky/slp3/}
\BIBentrySTDinterwordspacing

\bibitem{HIPAA2024}
HIPAA, ``Hipaa security rule nprm,'' \emph{HHS.gov}, Dec 2024.

\bibitem{obermeyer2019bias}
Z.~Obermeyer and S.~Mullainathan, ``Dissecting racial bias in an algorithm used to manage the health of populations,'' \emph{Science}, vol. 366, no. 6464, pp. 447--453, 2019.

\bibitem{cisco2024ai}
Cisco, ``Cisco ngo partner simprints to advance ethical, inclusive ai for face recognition biometrics,'' \url{https://blogs.cisco.com/our-corporate-purpose/cisco-ngo-partner-simprints-to-advance-ethical-inclusive-ai-for-face-recognition-biometrics}, Cisco, Tech. Rep., Jun 2024.

\bibitem{accenture2022ethicalai}
Accenture, ``Three considerations when building an intelligent data and ai platform,'' \url{https://www.accenture.com/nl-en/insights/artificial-intelligence/intelligent-data-ai-platorm}, Accenture, Tech. Rep., Apr 2022.

\bibitem{gartner2023ivr}
Gartner, ``Market guide for ivr systems and solutions,'' 2023.

\bibitem{Banham2018}
R.~Banham, ``Benefits and risks of automating customer service,'' \url{https://www.fm-magazine.com/issues/2018/dec/automating-customer-service-benefits-and-risks/}, Financial Management magazine, Tech. Rep., December 2018.

\bibitem{Owasp2010}
OWASP, ``Owasp secure coding practices - quick reference guide,'' \url{https://owasp.org/www-project-secure-coding-practices-quick-reference-guide/assets/docs/OWASP_SCP_Quick_Reference_Guide_v21.pdf}, 2010.

\bibitem{OWASPAI2025}
\BIBentryALTinterwordspacing
OWASP-AI-Exchange, ``Owasp – ai exchange,'' 2025. [Online]. Available: \url{https://owaspai.org/docs/}
\BIBentrySTDinterwordspacing

\bibitem{Clinton2025}
W.~Clinton and J.~Scott, ``Owasp ai security guidelines offer a supporting foundation for new uk government ai security guidelines, owasp top 10 for llm \& generative ai security,'' \url{https://genai.owasp.org/2025/01/31/owasp-ai-security-guidelines-offer-a-supporting-foundation-for-new-uk-government-ai-security-guidelines/}, Jan 2025.

\bibitem{Hao2019}
K.~Hao, ``This is how ai bias really happens and why it’s so hard to fix,'' \url{https://www.technologyreview.com/2021/12/03/1040618/problem-with-ai-training-data-bias/}, MIT Technology Review, Tech. Rep., Dec 2019.

\end{thebibliography}

\end{document}